\documentclass{aastex63}

\usepackage{xspace}
\usepackage{amsmath}
\usepackage{amstext}
\usepackage{amssymb}
\usepackage[amssymb]{SIunits}
\usepackage{soul}
\usepackage{hyperref}
\usepackage{cleveref}

\hypersetup{
	colorlinks = true,
	linkcolor = blue,
	citecolor = red
}

\begin{document}
\title{Comment on ``The Flyby Anomaly and the Gravitational-Magnetic Field Induced Frame-Dragging Effect around the Earth''}
\author{V. Guruprasad}
\affiliation{Inspired Research, New York, USA}

\begin{abstract}
\noindent
Independent radar data
	with larger discrepancies,
which were left out
	in JPL's 2008 summary on the anomaly,
rules out
	any actual change in motion,
relativistic or otherwise.


\end{abstract}
\section*{}

\citet{Mirza2019} presents
	a remarkable fit of
a conjectured enhanced frame-dragging induced by
	a coupling of gravitation and magnetism
to the ``flyby anomaly''
	as mainly described by
	\citet{Anderson2008},
and to its absence
	in later flybys.
The very inference of velocity and energy gains
	by \citet{Anderson2008}
		was fundamentally a mistake
because
	a basic significance of radar residuals
		in the context
	had been overlooked.
\citet{Antreasian1998} had disclosed
	kilometre-scale discrepancies in range data
		in NEAR's flyby
	against two coherent radars of
		the space surveillance network (SSN),
that were not examined by
	Anderson's team,
since they were too large to fit
	systematics or relativistic causes.
This is hardly surprising,
as even
	the Pioneer anomaly caught attention
only because of its similarity to
	the Hubble constant
	\citep{Nieto2007}.
However,
any actual change in motion to fit
	frame-dragging
or
	alternative hypotheses like dark matter
	\citep{Adler2009}
should have identically affected
	radar range,
so rather than disagree by two orders
	beyond the metre-scale accuracies of the two radars,
		the radar data should have concurred
			with the tracking.

Instead,
the radar residuals are quantitatively equivalent to
	an apparent doubling of the travel time of
		the telemetry tracking signal
	\citep{Prasad2015b},
that also fit
	NEAR's $\Delta V$
		to $1.3\%$
	and
	Rosetta's perigee advance of $0.34~\second$
		in its 2005 flyby
	\citep{Morley2006}
		to $2.3\%$
	\citep{Prasad2019a}.
We thus have an upper bound of $1\%$ on 
	frame-dragging or dark matter contributions
		if any.
The root mean square error and standard deviation
	given by \citet{Antreasian1998}
qualified them
	as $5\sigma$.
The fact that 
	the Doppler and range tracks
gave the same values for the anomalies
	seems to be why \citet{Anderson2008} assumed
		the $\Delta V$ was real,
but the equality signifies only that
	the signal comprising
		the carrier and modulation 
	was affected as a whole,
distinctly from
	radar signals on very similar paths.

To his credit,
\citet{Mirza2019} involves
	the field induction to explain
the absence of the anomalies
	in later flybys.
These absences more simply coincide with
	a switch from coherent to \emph{noncoherent}
		Doppler tracking
	after Cassini
	\citep{Chen2000,DeBoy2003},
however,
which led one of the co-authors (JKC) of
	\citet{Anderson2008}
to propose and guide the analysis
	reported in \citet{Prasad2019a}
		in the first place.
The absences in Rosetta's 2007 and 2009 flybys
	fit this signals explanation.
It has been suggested
	in deference to \citet{Mirza2019}
that the tracking signal and trajectory inconsistencies
	described in \citet{Prasad2019a}
must be ``fundamentally wrong''
	and should vanish if
		the trajectories are correctly recalculated,
given that
	the DSN Doppler design
	\citep{Moyer2000}
and
	Doppler tracking
	\citep{Kinman1992}
		are long established and should be correct.
While we agree that
	it would be the ideal case,
it needs to be pointed out that
	the trajectory reference and correlation
		in \citet{Prasad2019a}
	came \emph{entirely} from JPL Horizons
		\citep{JPLHorizons2015},
i.e., computed from
	the official trajectories 
		checked-in by NEAR and Rosetta teams
	and JPL's ephemerides.
Second,
JPL identified $\Delta V$
	as the velocity mismatch between
		trajectory segments
when extrapolated across
	a gap in the tracking
	\citep{Anderson2008},
so consistency of the trajectory was already
	in question.
Third,
NEAR came below
	the Hubble telescope's orbit
		at perigee,
and the tracking gaps in Galileo's flybys
	were below geostationary range:
atmospheric drag even made
	the anomaly uncertain in 1992
	\citep{Antreasian1998}.
Any dark matter or frame-dragging
	not already corrected for in
		JPL's trajectory software
should have affected
	the Hubble,
as well as
	the GPS and five hundred geosynchronous satellites
	by now.

As to the robustness of
	the DSN Doppler and the tracking technology,
the coherent transponders,
	which had all of the anomalies,
were qualified only on ground
	\citep{Mysoor1991},
and never even in low orbit
	unlike the noncoherent transponder
	\citep{Jensen2002}.
Lastly,
we show below
	an update of the radar residuals graphs
		from \citet{Prasad2019a}
highlighting
	the initial overlap of over a minute
		with DSN Goldstone.
As the trajectory is a direct fit to
	tracking data in a least-squares sense,
the overlap underscores that
	the discrepancy was not from extrapolation,
rather that
	the DSN data was off by almost a kilometre.
Even the five outliers from Millstone
	make sense
as geosynchronous satellites caught in its beam
	at the time.
Any further notion of
	a real causative force or relativistic effect
		would be wishful.
The only direction forward would be
	to reproduce the errant signal behaviour
		on ground.

\begin{figure}[h]
	\centering
	\includegraphics{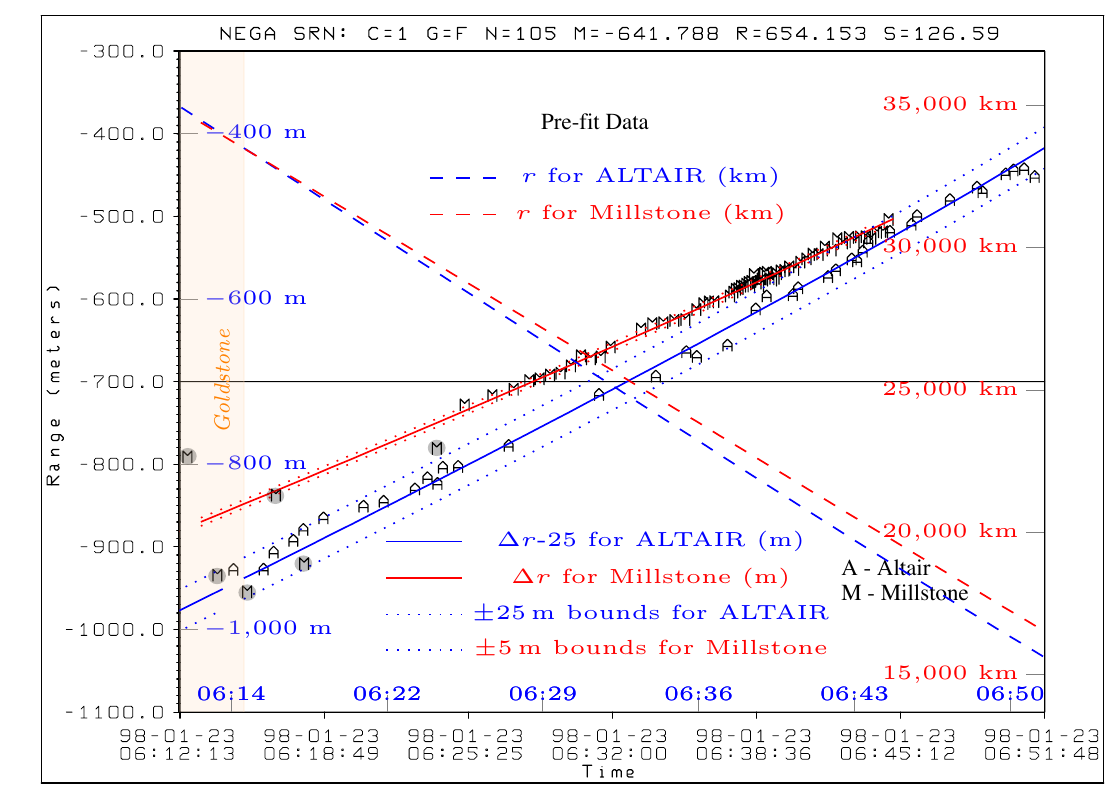}
	\caption{NEAR's SSN residuals with Goldstone overlap [updated from \citet{Prasad2019a}]}
\end{figure}



\begin{raggedright}
\bibliographystyle{apalike}


\end{raggedright}
\end{document}